\documentclass[10pt]{IEEEtran}
\IEEEoverridecommandlockouts
\usepackage{cite}
\usepackage{amsmath,amssymb,amsfonts}
\usepackage{algorithmic}
\usepackage{graphicx}
\usepackage{textcomp}
\def\BibTeX{{\rm B\kern-.05em{\sc i\kern-.025em b}\kern-.08em
    T\kern-.1667em\lower.7ex\hbox{E}\kern-.125emX}}

\newtheorem{theorem}{Theorem}[section]

\newtheorem{lemma}[theorem]{Lemma}

\newtheorem{definition}{Definition}
\usepackage{algorithm}

\usepackage{float}
\usepackage{tabularx}

\usepackage{url}

\usepackage[inline]{enumitem}
\usepackage{tabularx}
\usepackage{wrapfig,lipsum,booktabs}
\usepackage{pifont}
\usepackage{multirow}
\usepackage[table,xcdraw]{xcolor}
\usepackage{comment}
\usepackage{hyperref} 

\begin{document}
% \linespread{1.08075430507599209130405}
% \linespread{0.990135943771362}
% \linespread{0.963771362}

\title{\textit{dyGRASS}: Dynamic Spectral Graph Sparsification via Localized Random Walks on GPUs
}

% \author{

% \IEEEauthorblockN{Yihang Yuan}
% \IEEEauthorblockA{
% Stevens Institute of Technology\\
% yyuan22@stevens.edu}

% \and 
% \IEEEauthorblockN{Ali Aghdaei}
% \IEEEauthorblockA{
% University of California, San Diego\\
% aaghdaei@ucsd.edu}

% \and
% \IEEEauthorblockN{Zhuo Feng}
% \IEEEauthorblockA{
% Stevens Institute of Technology\\
% zfeng12@stevens.edu}
% }

\author{
\begin{tabular}{ccc}
Yihang Yuan & Ali Aghdaei & Zhuo Feng \\
Stevens Institute of Technology & University of California, San Diego & Stevens Institute of Technology \\
yyuan22@stevens.edu & aaghdaei@ucsd.edu & zfeng12@stevens.edu
\end{tabular}
}

\maketitle

\begin{abstract}

This work presents \textit{dyGRASS}, an efficient dynamic algorithm for spectral sparsification of large undirected graphs that undergo streaming edge insertions and deletions. At its core, \textit{dyGRASS} employs a random-walk-based method to efficiently estimate node-to-node distances in both the original graph (for decremental update) and its sparsifier (for incremental update). For incremental updates, \textit{dyGRASS} enables the identification of spectrally critical edges among the updates to capture the latest structural changes. For decremental updates, \textit{dyGRASS} facilitates the recovery of important edges from the original graph back into the sparsifier.
To further enhance computational efficiency, \textit{dyGRASS} employs a GPU-based non-backtracking random walk scheme that allows multiple walkers to operate simultaneously across various target updates. This parallelization significantly improves both the performance and scalability of the proposed \textit{dyGRASS} framework. Our comprehensive experimental evaluations reveal that \textit{dyGRASS} achieves approximately a 10$\times$ speedup compared to the state-of-the-art incremental sparsification (\textit{inGRASS}) algorithm while eliminating the setup overhead and improving solution quality in incremental spectral sparsification tasks. Moreover, \textit{dyGRASS} delivers high efficiency and superior solution quality for fully dynamic graph sparsification, accommodating both edge insertions and deletions across a diverse range of graph instances originating from integrated circuit simulations, finite element analysis, and social networks. We release our full implementation at: 
\href{https://github.com/Feng-Research/dyGRASS}{\texttt{\textcolor{blue}{https://github.com/Feng-Research/dyGRASS}}}
\end{abstract}

% \begin{IEEEkeywords}
% spectral graph theory, incremental sparsification, effective resistance, graph decomposition
% \end{IEEEkeywords}

% \input{2.Introduction}
\section{Introduction}
Graph-based methodologies play a fundamental role in addressing numerous challenges within electronic design automation (EDA), including logic synthesis \cite{stok2018eda3, 7293649}, verification \cite{hu2024deepic3}, layout optimization \cite{gao2021layout}, static timing analysis (STA) \cite{guo2022timing}, circuit partitioning \cite{karypis1997multilevel}, circuit modeling \cite{yang2022versatile}, and simulation \cite{hakhamaneshi2022pretraining}. Graph sparsification techniques, in particular, have been shown to be effective in accelerating a variety of compute-intensive EDA tasks, such as circuit simulations \cite{lengfei:tcad15,xueqian:tcad15, feng2016spectral,feng2020grass, liu2022pursuing}, allowing more efficient vector-less integrity verification of power grids \cite{zhang2022sf,zhiqiang:dac17}, and identifying worst-case on-chip temperature distributions \cite{zhao2022multilevel}.

Mathematics and theoretical computer science researchers have long studied the problem of simplifying large graphs through spectral graph theory \cite{batson2012twice,spielman2011spectral,Lee:2017,Chen2024BoostingGS}. Recent investigations into spectral graph sparsification have yielded significant advancements, allowing for constructing substantially sparser subgraphs while preserving the key graph spectral properties such as the first few eigenvalues and eigenvectors of the graph Laplacian. These breakthroughs have facilitated the development of nearly linear time algorithms for solving large sparse matrices and partial differential equations (PDEs) \cite{kyng2016approximate, chen2021multiscale}. Additionally, they have empowered various applications in graph-based semi-supervised learning (SSL), computing stationary distributions of Markov chains and personalized PageRank vectors, spectral graph partitioning, data clustering, max flow and multi-commodity flow of undirected graphs, and the implementation of nearly-linear time circuit simulation and verification algorithms \cite{miller:2010focs,spielman2011spectral,feng2016spectral,zhuo:dac18,zhao2022multilevel}.

While existing spectral graph sparsification methods are scalable for extracting sparsified graphs from scratch \cite{feng2016spectral,feng2020grass,fegrass, zhang2020sf,liu2022pursuing}, they are not suitable for handling cases with streaming edge insertions and deletions efficiently, which may pose a significant bottleneck in the modern EDA workflows. For example, the iterative nature of chip design means that even minor adjustments to the circuit schematic or layout, such as adding or removing circuit components, require the spectral sparsifiers to be updated with each change. In practical applications like power grid design and optimization \cite{Liu2023pGRASSS}, adding or removing a small number of metal wires requires reconstructing the entire spectral sparsifier, which quickly becomes computationally expensive. While recent theoretical advancements in dynamic graph sparsification algorithms offer promise \cite{kapralov2020fast,filtser2021graph,Abraham2016OnFD,Bernstein2020FullyDynamicGS}, the practical applicability of these techniques in real-world EDA workflows remains an open question. 
\begin{figure*}
    \centering
    \includegraphics[width=0.86598\textwidth]{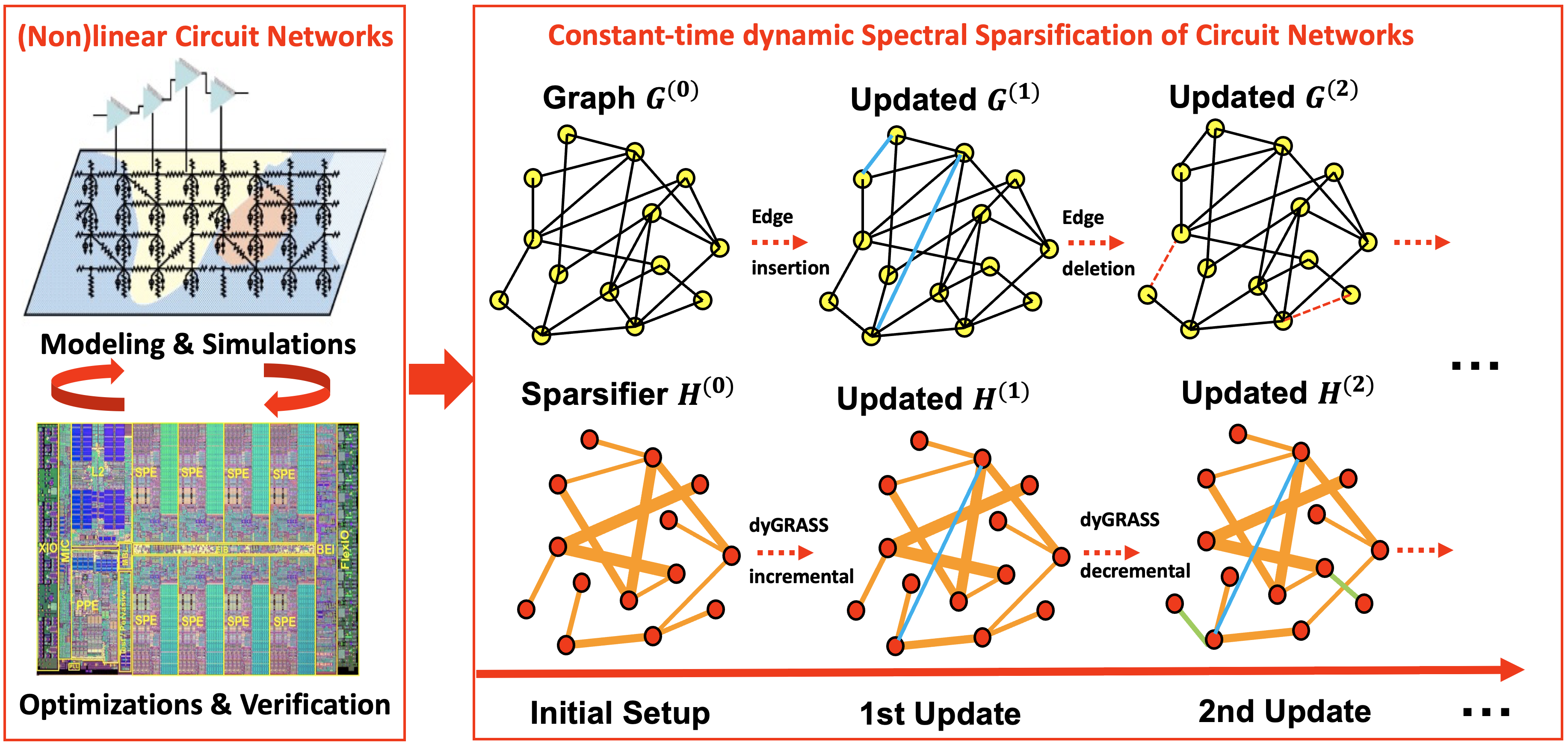}
    \caption{The proposed dynamic spectral sparsification method (\textit{dyGRASS}) and its applications in (non)linear circuit network modeling and optimization tasks. }
\label{fig:intro}
\end{figure*}

To address this gap, we propose a practically efficient algorithmic framework, \textit{dyGRASS}, for dynamic spectral sparsification of large undirected graphs. By leveraging localized random walks, \textit{dyGRASS} allows updating the spectral sparsifier in constant time for each edge insertion/deletion,  as shown in Fig. \ref{fig:intro}. In each edge update, \textit{dyGRASS} estimates the spectral distortion \cite{feng2016spectral,feng2020grass,zhang2020sf}  due to edge insertion/deletion by leveraging a localized random walk algorithm: for edge insertions \textit{dyGRASS} determines whether to retain or remove edges bounded by a resistance diameter, whereas for edge deletions \textit{dyGRASS} recovers edges along a selected path with minimum resistance to retain structural properties in the sparsifier. 

\begin{table}[]
\centering
\caption{Feature Comparison: \textit{inGRASS} vs. \textit{dyGRASS}}
\label{tab:feature_comparison}
\begin{tabular}{lcc}
\hline
\textbf{Aspect}      & \textbf{\textit{inGRASS}} \cite{aghdaei2024ingrass} & \textbf{\textit{dyGRASS}} \\ \hline
Setup Time                & \( O(n\log n) \) & N/A              \\ 
Update Time                & \( O(\log n) \) & \( O(1) \)              \\ 
Incremental Update          & \ding{51}        & \ding{51}        \\ 
Decremental Update          & \ding{55}        & \ding{51}        \\ 
Fully Dynamic         & \ding{55}        & \ding{51}        \\ 
GPU Acceleration      & \ding{55}        & \ding{51}        \\ \hline
\end{tabular}
\end{table}

As shown in Table \ref{tab:feature_comparison} where $n$ denotes the number of vertices in the graph, the state-of-the-art \textit{inGRASS} framework requires $O(n \log n)$ time for the setup phase and supports only incremental edge insertions \cite{aghdaei2024ingrass}, whereas \textit{dyGRASS} eliminates the costly setup phase and process each edge insertion or deletion in constant time, offering a more efficient and comprehensive approach to dynamic spectral graph sparsification. Moreover, unlike \textit{inGRASS} that keeps using the resistance embedding of the initial sparsifier obtained during the setup phase for processing newly added edges, \textit{dyGRASS} offers a fully dynamic sparsification framework for continuously exploiting the latest graph structure for achieving a nearly-optimal sparsification solution. In addition, our \textit{dyGRASS} framework allows for utilizing GPUs to handle large-scale updates efficiently, achieving substantial speedups for processing large  dynamically  evolving graphs. The key contributions of this work have been summarized as follows:
\begin{enumerate}
\item  We propose \textit{dyGRASS} for dynamically updating spectral graph sparsifiers. Unlike the latest \textit{inGRASS} framework \cite{aghdaei2024ingrass} that requires an expensive setup phase and supports only a limited number of incremental edge insertions, \textit{dyGRASS} can handle any size of edge insertions/deletions with greatly improved efficiency.

\item The proposed \textit{dyGRASS} framework leverages localized random walks to efficiently approximate graph distances and assess spectral distortion caused by each update. Since \textit{dyGRASS} is a strongly local algorithm, it enjoys a constant time complexity for each edge insertion/deletion operation. 

\item By accelerating random walks on Graphics Processing Units (GPUs), \textit{dyGRASS} achieves a   $70\times$  runtime speedup over the CPU-based implementation and at most $10\times$    speedup over the state-of-the-art method \cite{aghdaei2024ingrass} on real-world large graphs, while preserving solution quality. 
\end{enumerate}

\section{Background}\label{sec:background}
\subsection{Graph Laplacian, quadratic form, and spectral similarity}

For a weighted undirected graph $G = (V, E, w)$,   $V$ ($|V| = n$) and $E$ ($|E| = m$) represent the sets of vertices and edges, respectively, and $w$ denotes a positive weight function with $w_{i,j}$ denoting the edge  weight. Let $A$ denote the $n \times n$ adjacency matrix, then the Laplacian matrix $L$ is defined as $L:= D - A$, where $D$ is the degree matrix. The eigenvalues and eigenvectors of $L$ offer insights into graph structure properties such as the number of connected components and vertex connectivity. The Laplacian quadratic form $x^{\top}L x$ is a fundamental tool in spectral graph theory, enabling analysis of graph cuts, clustering, and conductance.

Spectral similarity between the Laplacians of graphs $G$ and $H$ can be quantified using the following inequality \cite{spielman2008graph}:

\begin{equation}
\frac{x^{\top}L_G x}{\epsilon} \leq x^{\top}L_H x \leq \epsilon x^{\top}L_G x,
\end{equation}
where $L_G$ and $L_H$ represent the Laplacian matrices of $G$ and $H$, with a parameter $\epsilon > 0$. A smaller $\epsilon$ or relative condition number $\kappa (L_G, L_H)$ indicates greater spectral similarity between $G$ and $H$.
\subsection{Spectral sparsification}
Spectral sparsification methods are used to represent a large graph in a more concise form by removing redundant or less important edges while preserving its essential structural and spectral properties. Given a graph $G = (V, E, w)$ with its Laplacian matrix $L_G$, its sparsifier $H = (V, E', w')$ is its subgraph with its Laplacian matrix $L_H$, where $|E'| \ll |E|$, while preserving the original spectral graph properties, such as the Laplacian eigenvalues and eigenvectors. Spectral sparsification retains these properties, enabling efficient algorithms and analyses while reducing computational costs. Spectral sparsification has been applied to efficiently solve large systems of equations and tackle graph-based computational problems. For example, instead of directly solving the original Laplacian matrix of graph $G$, the Laplacian of the sparsified graph $H$  can be used as a preconditioner in standard iterative sparse matrix algorithms for reducing total solution time \cite{feng2020grass}.

\subsection{Effective resistance}

\begin{definition} \label{def:ER}
  The effective resistance between vertices   $(p, q) \in V$ is defined as
\begin{equation}\label{eq:eff_resist0}
    R^{\text{eff}}_{p,q} := b_{p,q}^\top L^\dagger b_{p,q}= \sum\limits_{i= 2}^{|V|} \frac{(u_i^\top b_{p,q})^2}{u_i^\top L u_i},
\end{equation}
where $L^{\dagger}$ denotes the Moore-Penrose pseudo-inverse of the graph Laplacian matrix  $L$,   $u_{i} \in \mathbb{R}^{|V|}$ for $i=1,...,|V|$ denote the  unit-length, mutually-orthogonal  eigenvectors corresponding to  Laplacian eigenvalues $\lambda_i$ for $i=1,...,|V|$,     
${b_{p}} \in \mathbb{R}^{|V|}$ denotes the standard basis vector with all zero entries, with the $p$-th entry equal to 1, and ${b_{p,q}}=b_p-b_q$.
\end{definition}

Computing effective resistances directly according to Definition \ref{def:ER} proves impractical for large graphs. However, our dynamic algorithm only requires fast estimations of edge effective resistances, which can be achieved using a highly scalable, constant-time   random walk approach.

\subsection{Random walk}

A random walk on a graph is a stochastic process where a walker starts at a vertex and moves to a randomly chosen neighbor at each step. Random walks are a fundamental tool in graph analysis, offering insights into structural and spectral properties and enabling sampling, distance estimation, and connectivity analysis without full graph exploration \cite{Wang2015DeterministicRW}. 
% Let $G = (V, E)$ be an undirected graph, with $|V| = n$ vertices and $|E| = m$ edges. 
For any two vertices $p, q \in V$, the \textbf{hitting time} $h_{p,q}$ is the expected number of steps for a walker starting at $p$ to reach $q$ for the first time. 

\begin{lemma}
For any two vertices $p$ and $q$ in $G$, the \textbf{commute time} $C_{p,q}$ is the expected round-trip time between $p$ and $q$, which is directly proportional to the effective resistance distance between $p$ and $q$:
$C_{p,q} = h_{p,q} + h_{q,p} = 2m \cdot R^{\text{eff}}_{p,q}$ \cite{Chandra1996}.
\end{lemma}

The above relationship underscores the connection between random walk dynamics and graph spectral properties, which allows for an efficient approximation of effective resistance using random walk metrics.

 However, using a Monte Carlo approach to approximate commute times directly remains impractical for large graphs, as the computational effort scales with the total edge count $m$.

 Recent work by Peng et al.~\cite{peng2021local} proposed local algorithms for estimating effective resistance by approximating random walk commute times and Laplacian pseudo-inverses. Their approach, such as EstEff-TranProb, uses a series expansion of the transition matrix and performs multiple $\ell$-step random walks to estimate multi-hop transition probabilities. While theoretically elegant and suitable for graphs with small mixing time, the required number of steps $\ell$ grows quickly as the spectral gap closes. In graphs with high spectral radius, this makes the method impractical for large graphs due to the need for long-range random walks. Moreover, the reliance on long-range walks makes them costly for online or streaming settings. In contrast, our method avoids dependence on global mixing properties or pseudo-inverse approximations. By leveraging localized \emph{non-backtracking random walks}, we construct efficient upper bounds on effective resistance, enabling fast edge filtering with minimal exploration. This makes our approach more robust in the face of slow mixing and more suitable for dynamic, large-scale graphs.

\subsection{Dynamic graph algorithms}
Dynamic graph algorithms are designed to efficiently handle graphs that undergo updates over time. These algorithms can be categorized as follows:

\begin{enumerate}
    \item Incremental algorithms: These algorithms efficiently handle additions of edges to the graph. They update the data structure representing the graph to incorporate the new elements while minimizing the computational cost.

    \item Decremental algorithms: These algorithms efficiently handle edge deletions from the graph. They update the data structure to reflect the removal of elements while preserving the integrity of the graph.

    \item Fully dynamic algorithms: These algorithms handle both additions and deletions of edges from the graph. They dynamically adjust the data structure to accommodate changes while maintaining efficient query performance.
\end{enumerate}

This work focuses on an efficient yet fully dynamic graph update algorithm, leveraging a localized random-walk-based distance estimation technique to manage edge insertions and deletions within the spectral graph sparsifier.

\section{\textit{dyGRASS}: Dynamic Graph Sparsification}\label{sec:alg}
\subsection{Overview of \textit{dyGRASS}}
\begin{figure}
    \centering
    \includegraphics[width=0.48\textwidth]{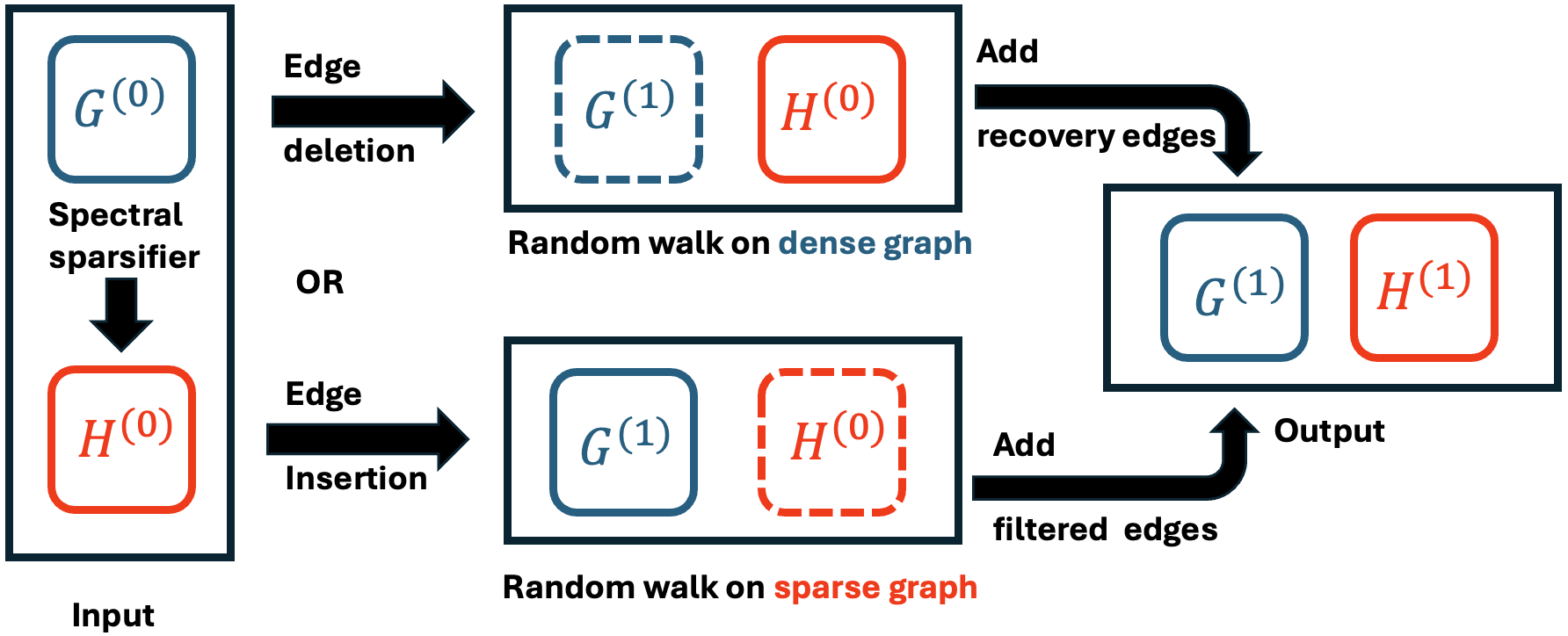}
    \caption{Overview of the \textit{dyGRASS} framework, with dashed lines indicating the graph where localized random walks are initialized.}
\label{figure:dygrass_overview}
\end{figure}

The proposed \textit{dyGRASS} algorithm dynamically updates the spectral sparsifier in response to streaming edge updates (insertions and deletions), as illustrated in Fig.\ref{figure:dygrass_overview}. The algorithm takes as input the original graph $G^{(0)}$ and its initial spectral sparsifier $H^{(0)}$, which is precomputed using the \textit{GRASS} method \cite{feng2020grass}. Our approach comprises two main phases: (1) incremental spectral graph sparsification and (2) decremental spectral graph sparsification. By  integrating these two phases, the \textit{dyGRASS} algorithm efficiently updates the initial spectral sparsifier $H^{(0)}$ without recomputing the sparsifier from scratch, thereby significantly reducing computational overhead.

\subsection{Incremental spectral sparsification with \textit{dyGRASS}}

\textbf {Spectral Impact of Edge Insertions:} Suppose an edge $e_{p,q} \notin E^{(l)}$ is newly introduced between vertices $p$, $q$ during the $l$-th iteration, updating the original graph $G^{(l)} = (V, E^{(l)}, w^{(l)})$. When incorporating $e_{p,q}$ into current sparsifier $H^{(l)} = (V, E'^{(l)}, w'^{(l)})$, the eigenvalue perturbation for each eigenvector of $H^{(l)}$ can be assessed by the following lemma \cite{zhang2020sf}.
\begin{lemma}
Let $H = (V, E', w')$, where $w':E' \rightarrow \mathbb{R}+$, denote the sparsified weighted graph of $G$, and $L_H$ denotes its Laplacian matrix. The $i$-th Laplacian eigenvalue perturbation $\delta \lambda_{i}$ due to $\delta L_H=  w'_{p,q}b_{pq} b^\top_{pq}$ can be computed as:
\begin{equation}
\delta \lambda_{i} =  w'_{p,q} (u_{i}^{\top} b_{pq})^2,
\end{equation}
where $u_{i}$ represents the eigenvector corresponding to the $i$-th eigenvalue $\lambda_{i}$ of the Laplacian matrix $L_H$.
\end{lemma}

Thus, if $e_{q,p}$ has a larger $w'_{p,q} (u_{i}^{\top} b_{pq})^2$, it is considered spectrally critical to $\lambda_i$. This indicates that including this edge in the updated sparsifier would cause a larger perturbation to the $i$-th eigenvalue $\lambda_i$ and its corresponding eigenvector $u_i$.

\begin{lemma}
Construct a  weighted eigensubspace matrix $U_K$ for $K$-dimensional spectral graph embedding using the first $K$  Laplacian eigenvectors and eigenvalues as follows:
\begin{equation}\label{subspace}
U_K=\left[\frac{u_2}{\sqrt {\lambda_2}},..., \frac{u_K}{\sqrt {\lambda_K}}\right], 
\end{equation}
then  the spectral distortion of the new edge $e_{p,q}$ will become the total  $K$-eigenvalue  perturbation $\Delta_K$ when $K \rightarrow N $\cite{zhang2020sf}:
\begin{equation}\label{eq:delta}
\Delta_K := \sum\limits_{i = 2}^{{K}}  \frac{\delta { {\lambda}_{i}^{}}}{\lambda_i} =  w_{p,q} \|U_K^\top b_{pq}\|^2_2 \approx w_{p,q} R^{\text{eff}}_{p,q}.
\end{equation}
\end{lemma}

Lemma 3.2 defines the spectral distortion $\Delta_K$ of an edge as the product of its edge weight and effective resistance between its two end vertices. We next show that the \emph{all} distortion
remains bounded by the condition number $\kappa(L_G, L_H)$.

For a graph $G$ and a subgraph $H$ (obtained by deleting edges), consider the generalized eigenvalue problem for the matrix pencil 
$(L_H, L_G)$ and apply the Courant-Fischer theorem. Specifically,
we have:
\begin{equation}\label{eq:cf1}
\max_{\substack{p,q \in V \\ p \neq q}}
\frac{b_{p,q}^\top L_H^+\,b_{p,q}}
     {b_{p,q}^\top L_G^+\,b_{p,q}}
\;\;\le\;\;
\max_{\substack{\|v\|\neq0 \\ v^\top \mathbf{1} = 0}}
\frac{v^\top L_H^+\,v}{v^\top L_G^+\,v}
\;=\;
\lambda_{\max},
\end{equation}
where 
% $b_{p,q}$ is the incidence vector of edge $e_{p,q}$, and 
$\mathbf{1}$ is the all-one vector. 
\begin{equation}\label{eq:cf2}
\frac{b_{p,q}^\top L_H^+\,b_{p,q}}
     {b_{p,q}^\top L_G^+\,b_{p,q}}
\;=\;
\frac{R^{\text{eff}}_{p,q,H}}{R^{\text{eff}}_{p,q,G}},
\end{equation}
From equations~\eqref{eq:cf1} and~\eqref{eq:cf2}, we see that taking the maximum over all node pairs $(p,q)$ 
implies
\begin{equation}\label{eq:maxratio}
\max_{p,q}\,\frac{R^{\text{eff}}_{p,q,H}}{R^{\text{eff}}_{p,q,G}}
\;\;\le\;\;
\lambda_{\max}(L_G\,L_H^+).
\end{equation}
\medskip

Furthermore, to find $\lambda_{\min}(L_G,L_H)$, one applies the 
Courant-Fischer theorem again:
\begin{equation}
1 \;\le\; \lambda_{\min}
=\;
\min_{\substack{\|v\|\neq0 \\ v^\top \mathbf{1}=0}}
\frac{v^\top L_G\,v}{v^\top L_H\,v}
\;\le\;
\min_{\substack{\|v\|\neq0 \\ v(i)\in\{0,1\}}}
\frac{v^\top L_G\,v}{v^\top L_H\,v}.
\end{equation}
Restricting $v$ to $\{0,1\}$-valued entries treats $v$ as a cut 
and sums up the boundary edges in $G$ vs.\ $H$.  
In particular, if there exists a vertex $u$ of degree 1 in $G$ and 
the same edge connecting $u$ to its neighbor 
is preserved (with the same weight) in $H$, 
then assigning $v(u)=1$ and $v(i)=0$ for $i\neq u$ yields
\begin{equation}
\frac{v^\top L_G\,v}{v^\top L_H\,v}
\;=\;\frac{\omega_{u,\text{neighbor}}}{\omega_{u,\text{neighbor}}}
\;=\;1,
\end{equation}
establishing $\lambda_{\min}(L_G,L_H)\le1$.  
Since a leaf vertex cannot become isolated in $H$ 
without invalidating the sparsifier, 
we also have $\lambda_{\min}(L_G,L_H)\ge1$. 
Hence $\lambda_{\min}(L_G,L_H)=1$.  
If $L_G L_H^+$ thus has its smallest eigenvalue equal to 1, 
then $\lambda_{\max}(L_G L_H^+)$ is precisely the relative 
condition number $\kappa(L_G,L_H)$.  
Consequently,
\begin{equation}
\omega_{p,q}\,R^{\text{eff}}_{p,q,H}
\;\le\;
\frac{R^{\text{eff}}_{p,q,H}}{R^{\text{eff}}_{p,q,G}}
\;\le\;
\kappa(L_G,L_H),
\quad
\forall\,p,q\in V.
\end{equation}
In other words, no single edge's effective-resistance ratio 
can exceed $\kappa(L_G,L_H)$, bounding the spectral distortion 
by the condition number.

Equation~\eqref{eq:delta} indicates that an edge $e_{p,q}$ with a large product $w_{p,q}\,R^{\text{eff}}_{p,q}$ can impose significant spectral distortion $\Delta_K$. Yet, \emph{directly computing} the $R^{\text{eff}}_{p,q}$ for every inserted edge is prohibitively expensive on large graphs. 
To circumvent this bottleneck, we adopt a \emph{non-backtracking random walk} approach that efficiently \emph{approximates} or upper-bounds $R^{\text{eff}}_{p,q}$ to enable rapid edge filtering.

\textbf{Non-Backtracking Random Walks and Approximate Resistance Distance:} 
At the core of our method lies the observation that the accumulated effective resistance along the shortest path between two vertices $p$ and $q$ provides an upper bound on the true $R^{\text{eff}}_{p,q}$. Specifically, if one could identify the shortest path $P^*$ from $p$ to $q$, then the sum $\sum _{e \in P^*} r_e$ (where $r_e = 1/w_e$) would serve as a valid upper bound on $R^{\text{eff}}_{p,q}$, as resistance in parallel can only reduce total resistance.

However, directly computing shortest paths or all-pairs effective resistance remains computationally prohibitive in large-scale graphs. To circumvent this, we adopt a non-backtracking random walk (NBRW) approach, which probabilistically explores paths from $p$ without immediately revisiting the previous node, emulating a depth-first search pattern but in a randomized and lightweight manner. On a tree-structured graph, the path generated by NBRW from $p$ to $q$ corresponds exactly to the unique shortest path, making the accumulated resistance along the walk equal to $R^{\text{eff}}_{p,q}$.

In general graphs, especially those sparsified using low-stretch spanning trees with off-tree edges, the NBRW path may not be the shortest. Nonetheless, the resistance accumulated along the walk still serves as a valid upper bound on the effective resistance, and—crucially—shorter paths are more likely to be sampled by repeated NBRWs. Thus, by initiating multiple non-backtracking walks from $p$, we estimate the resistance via:
\begin{equation}
\tilde{R}_{p,q} := \min_{1 \le i \le s} \left( \sum_{e \in P^{(i)}} r_e \right)
\end{equation}
where each $P^{(i)}$ is a sampled non-backtracking path reaching $q$ and $s$ is the number of walkers.

This Monte Carlo strategy enables a practical, localized approximation of the effective resistance. The approximate distortion is then defined as $\tilde{\Delta} := \omega_{p,q}\tilde{R}_{p,q}$, and a distortion threshold $K$ is imposed to guide the random walk: walkers terminate their search once the cumulative approximate distortion exceeds this threshold ($\omega_{p,q} \tilde{R}_{\text{cumulative}} > K$), dynamically adjusting their search range rather than using a fixed step limit.
In other words, the threshold $K$ acts as an effective proxy to filter edges in the sparsification process. For a newly introduced edge $e_{p,q}$:

\begin{itemize}
\item If vertex $q$ is not reachable from vertex $p$ within the distortion threshold $K$, the edge $e_{p,q}$ is considered spectrally critical and should be included in the sparsifier.
\item If vertex $q$ is reachable within the distortion threshold $K$, the edge $e_{p,q}$ can be safely pruned without significantly impacting spectral similarity.
\end{itemize}

Moreover, the threshold $K$ provides a tunable tradeoff between sparsifier density and spectral accuracy:
\begin{itemize}
    \item Smaller $K$ results in more edges being retained, lowering the condition number but increasing sparsifier density;
    \item Larger $K$ leads to more aggressive pruning, reducing density but potentially increasing the condition number.
\end{itemize}

To account for randomness and prevent false negatives (i.e., failing to reach $q$ despite it being within range), \textit{dyGRASS} deploys $s$ independent walkers from $p$. The probability of all walkers missing $q$ decreases exponentially with $s$. 

While the distortion threshold $K$ controls search adaptively, it does not bound the maximum number of steps. In regions where edge weights are large (i.e., resistances are small), many steps may be required before the accumulated distortion exceeds $K$. In fact, the worst-case number of steps can be loosely bounded by: $\frac{K}{\omega_{\min} / \omega_{\max}}$, where $\omega_{\min}$ is the minimum weight among candidate inserted edges, and $\omega_{\max}$ is the maximum edge weight in the graph. This implies that, even for moderate $K$, a walk might traverse a large portion of the graph in low-resistance regions.

To avoid such unbounded exploration and ensure computational efficiency, we introduce a fixed \emph{maximum step limit} $T$ for each walker. Thus, a walker in \textit{dyGRASS} may terminate under three conditions:
\begin{itemize}
    \item The cumulative distortion $\omega_{p,q}\tilde{R}_{\text{cumulative}}$ exceeds $K$,
    \item The number of steps reaches the maximum threshold $T$,
    \item The walker enters a leaf node (i.e., has no further neighbors to explore).
\end{itemize}

Importantly, when a walker terminates due to step limit $T$ without reaching $q$, \textit{dyGRASS} conservatively assumes that $q$ lies outside the search region, and the edge $e_{p,q}$ is retained in the sparsifier. Although this may slightly increase sparsifier density, it ensures that no spectrally critical edges are erroneously removed—thus preserving or improving the spectral condition number.

$T$ effectively acts as a hyper-parameter in \textit{dyGRASS}; it does \emph{not} scale with graph size. Instead, $T$ controls how long each non-backtracking walker explores before terminating. A larger $T$ can accommodate a higher distortion threshold $K$, promoting more aggressive edge pruning (and potentially yielding a sparser graph) but risking an increase in the condition number. Conversely, a smaller $T$ strictly caps random-walk lengths and helps constrain the condition number to remain low, though it may retain additional edges.

The step limit $T$ caps the computational cost of each walk to at most $T$ steps, and with $s$ walkers per edge update, the total work is bounded by $s \times T$, which is a constant and independent of the graph size. This guarantees that \textit{dyGRASS} achieves \emph{constant-time} complexity $\mathcal{O}(1)$ per edge update in practice.

In our experiments, we fix $T=100$ as a reasonable default. This choice reflects practical observations that extremely high condition numbers are uncommon in many real-world EDA and network graphs, so a moderate $T$ sufficiently captures local structure information without excessive search overhead. Users can adjust $T$ based on their desired tradeoff between pruning aggressiveness (governed by $K$) and runtime constraints, all while preserving the $\mathcal{O}(1)$ update complexity characteristic of \textit{dyGRASS}.

By combining non-backtracking random walks, a distortion threshold $K$, and a capped step limit $T$, \textit{dyGRASS} achieves a practical approximation of effective resistance distances in large graphs. This framework avoids the computational cost of exact shortest-path or pseudo-inverse methods, preserving high spectral fidelity at minimal overhead.

\subsection{Decremental spectral graph sparsification}
Consider an edge $e_{p,q} \in E$ to be deleted from the original graph $G^{(0)}$. The goal is to efficiently update $H^{(0)}$ while preserving the spectral similarity between $G$ and $H$. Our decremental method will first determine if $e_{p,q}$ belongs to the $H^{(0)} = (V, E'^{(0)}, w'^{(0)})$:

\textbf{if $\mathbf{e_{p,q} \notin E'^{(0)}}$:} $H^{(0)}$ remains unchanged, since $e_{p,q}$ is not part of the sparsifier.

\textbf{if $\mathbf{e_{p,q} \in E'^{(0)}}$:} The sparsifier must be updated by recovering critical edge(s) from $G^{(1)}$ to $H^{(0)}$, so as to preserve crucial distances in the new graph.

\smallskip
\noindent
\textbf{Distance estimation using non-backtracking random walks:}
Let us assume that $e_{p,q} \in E$ is deleted from $G^{(0)}$ and also exists in the spectral sparsifier $H^{(0)}$. Simply removing this edge from $H^{(0)}$ may isolate certain vertices or significantly perturb their distances. In other words, deleting a sparsifier edge can disrupt connectivity and inflate effective resistances between previously connected vertices. 

To address these challenges, it is necessary to \emph{recover edges} from the updated original graph $G^{(1)}$ (after the deletion) and add them back into $H^{(0)}$. We achieve this by performing localized NBRWs:

\begin{itemize}
    \item \textbf{Multiple Walkers with Step Limit $T$:} For the deleted edge $e_{p,q}$, we launch several non-backtracking walkers from $p$ (or from $q$). Each walker proceeds up to $T$ steps without traversing the same edge in consecutive steps. 
    \item \textbf{Accumulated Resistance:} As a walker moves along a path $P$, it accumulates the sum of resistances ($1/w_e$) for the edges $e \in P$. This sum represents a \emph{distance} measure in the updated graph $G^{(1)}$. 
    \item \textbf{Selecting the Minimum-Resistance Path:} Among all paths that successfully reach $q$, only the path with the \emph{smallest} total resistance is chosen for recovery. The edges along this path are added to the sparsifier to restore the updated resistance distances between $p$ and $q$. 
\end{itemize}

In the event that \emph{no} walker successfully reaches $q$ within $T$ steps, it implies that the resistance distance between $p$ and $q$ after edge deletion is sufficiently large and no specific edge recovery is required to maintain spectral fidelity. \textit{dyGRASS} still recovers a few local edges near $p$ and $q$ to prevent the creation of isolated components from forming in the sparsifier. 
This approach ensures the updated sparsifier reflects the structural changes in $G^{(1)}$, preserving spectral integrity without exhaustive searches. This localized NBRW-based scheme ensures the structural and spectral integrity of the updated sparsifier when handling edge deletions. By bounding each walker’s path length to $T$, the time spent exploring is limited, yielding an efficient update procedure that seamlessly complements \textit{dyGRASS}’s incremental phase.

\subsection{The \textit{dyGRASS} algorithm flow} 

\begin{algorithm}[h!]
\small
\caption{\textit{dyGRASS}: Dynamic Spectral Sparsification}
\label{alg:decremental}
\begin{algorithmic}[1]
\STATE \textbf{Input:} 
$G^{(0)}$: Original graph, 
$H^{(0)}$: Sparsifier, 
$\text{EdgeStream}$: Stream of updates, 
distortion threshold $K$,
maximum step size $T$, 
number of walkers $s$, 
update type $choice$
\STATE \textbf{Output:} $H^{(t)}$: Updated sparsifier
\STATE Initialize $H^{(t)} \gets H^{(0)}$

\FOR{each update $(\text{type}, (u, v))$ in $\text{EdgeStream}$}
    \IF{choice == \textbf{``insertion"}}
        \STATE 1. Add $(u, v)$ to $G^{(t)} \to G^{(t+1)}$
        \STATE 2. Perform non-backtracking random walks from $u$ to $v$ on $H^{(t)}$, using $K$, $T$, and $s$
        \IF{$v$ not reachable by any walker}
            \STATE \textbf{then} add $(u, v)$ to $H^{(t)}$
        \ENDIF
    \ELSIF{choice == \textbf{``deletion"}}
        \STATE Remove $(u, v)$ from $G^{(t)} \to G^{(t+1)}$
        \IF{$(u, v) \in H^{(t)}$}
            \STATE 1. Remove $(u, v)$ from $H^{(t)}$
            \STATE 2. Perform non-backtracking random walks between $u$ and $v$ on $G^{(t+1)}$, up to step limit $T$
            \STATE 3. Select a path with minimum accumulated resistance in $G^{(t+1)}$
            \STATE 4. Recover edges from this path into $H^{(t)}$
        \ENDIF
    \ENDIF 
    \STATE Update $H^{(t)} \to H^{(t+1)}$
\ENDFOR
\RETURN $H^{(t)}$
\end{algorithmic}
\end{algorithm}
The  flow of the proposed dynamic graph sparsification algorithm (\textit{dyGRASS}) is shown in Algorithm \ref{alg:decremental} and described as follows:\\
% Explanation of lines
\textbf{Lines 1--3:} The inputs include the original graph $G^{(0)}$ and its initial spectral sparsifier $H^{(0)} $ precomputed using \textit{GRASS} method \cite{feng2020grass}, and a stream of edge updates (EdgeStream). The threshold $K$ and step limit $T$ define local exploration for each walker, and $s$ denotes the number of walkers.
\\
\textbf{Lines 5--9:} For edge insertions, \textit{dyGRASS} adds $e_{u,v}$ to $G^{(t)}$ and checks via NBRW in $H^{(t)}$. if $(u,v)$ is unreachable within $K$ or step $T$, it is considered critical and is kept; otherwise, it can be pruned. \\
\textbf{Lines 12--17:} For a deletion, \textit{dyGRASS} removes $e_{u,v}$ from $G^{(t)}$. if $e_{u,v}$ is also in $H^{(t)}$, it is removed and a minimum resistance path is recovered via random walks in $G^{(t+1)}$.. If no path is found, adjacent edges near $(u,v)$ will be recovered to prevent forming isolated components.

The per-update time complexity of \textit{dyGRASS} is effective $O(1)$, under fixed parameters $T$ and  $s$, independent of graph size.

\subsection{A Case Study for Dynamic Spectral Graph Sparsification}

\begin{figure}[t]
    \centering
    \includegraphics[width=0.50\textwidth]{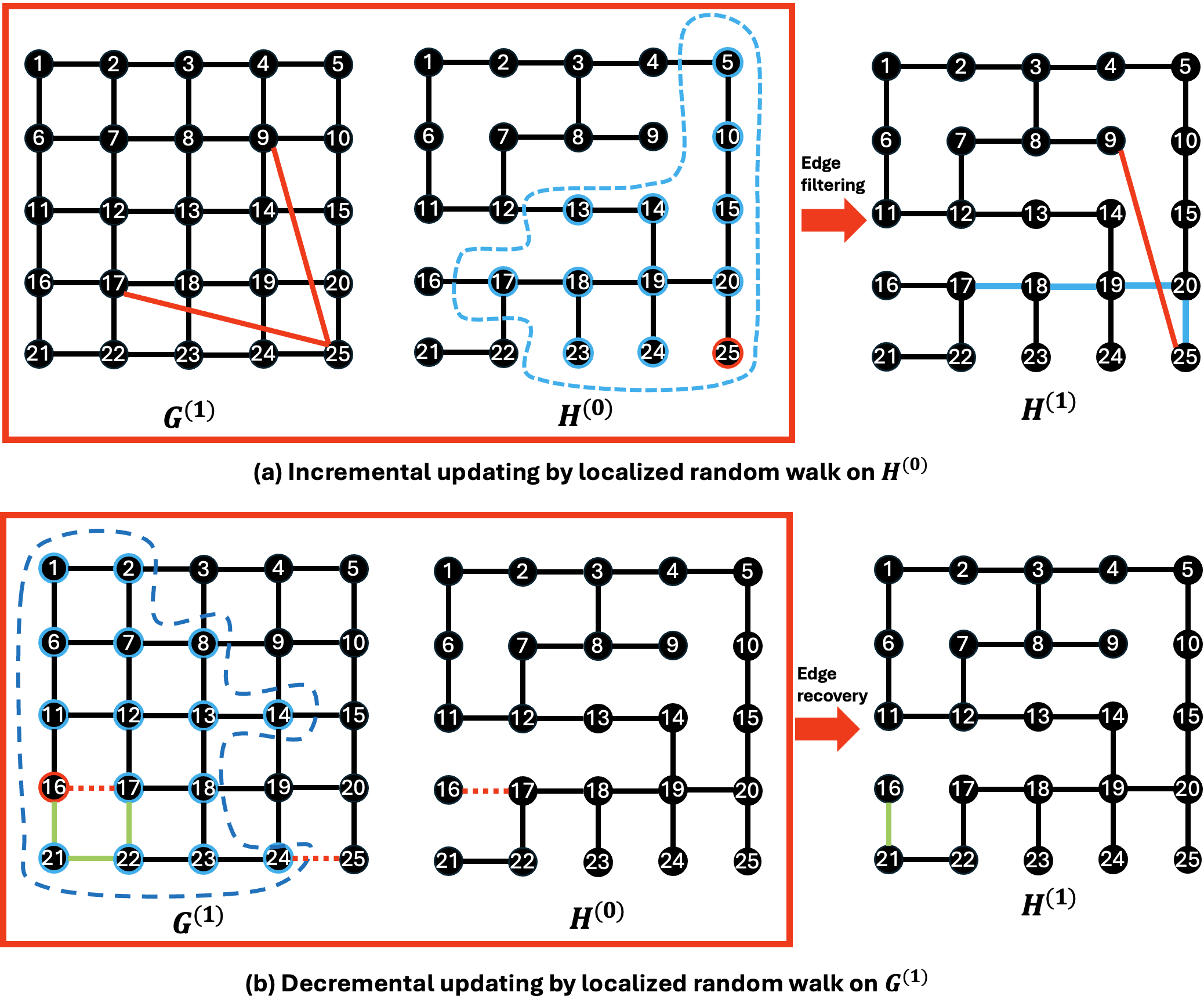}
    \caption{Schematic of the \textit{dyGRASS} algorithm illustrating incremental updates of $e_{25,17}$ and $e_{25,9}$, as well as decremental updates of $e_{16,17}$ and $e_{24,25}$.}
    \label{figure:inc_up}
\end{figure}

Fig.~\ref{figure:inc_up} presents an example of \textit{dyGRASS} managing both incremental and decremental updates on a simple undirected graph. In Fig.~\ref{figure:inc_up}(a), two edges, $(25,17)$ and $(25,9)$, are introduced in the original graph $G^{(0)}$, producing the updated graph $G^{(1)}$. Non-backtracking random walkers begin at vertex~25 in the current sparsifier $H^{(0)}$, using a distortion threshold $K=4$ (equivalent to a maximum walk length of four steps), as illustrated by the blue dashed circle.

From these walks, vertex~9 remains unreachable, indicating that $(25,9)$ is spectrally critical that must be included in the updated sparsifier. Conversely, since vertex~17 is reached by at least one walker, the newly introduced edge $(25,17)$ is excluded from the sparsifier. The resulting sparsifier is denoted as $H^{(1)}$.

In Fig.~\ref{figure:inc_up}(b), two edges, $(16,17)$ and $(24,25)$, are deleted from $G^{(0)}$, producing another updated graph $G^{(1)}$. The sparsifier after edge deletion but before recovery is shown as $H^{(0)}$. Since $(24,25)$ is \emph{not} in $H^{(0)}$, no further action is required. However, the deleted edge $(16,17)$ is in $H^{(0)}$, prompting a sparsifier update. Walkers are thus initialized from vertex~16 in $G^{(1)}$, each limited to four steps. Among the discovered paths, the one with the minimum accumulated resistance distance includes $(16,21)$, $(21,22)$, and $(22,17)$ (highlighted in green). Only $(16,21)$ is not already in $H^{(0)}$ and is therefore recovered into $H$. The updated sparsifier $H^{(1)}$ is then obtained.

This example illustrates how \textit{dyGRASS} dynamically identifies and preserves spectrally significant edges under incremental and decremental operations, maintaining strong spectral similarity with the evolving original graph.

\subsection{GPU-accelerated random walks}
Although \textit{dyGRASS} maintains an $O(1)$ update time per edge in theory, the actual constant factors can become significant if the step limit $T$ or the number of walkers $s$ grows large. For incremental updates, a high condition number may necessitate an increased $T$ to correctly capture high-distortion edges. In decremental updates, despite the original graph often having a smaller radius, deploying additional walkers can still increase the total workload. Consequently, each update can involve a substantial amount of random-walk exploration when $T$ or $s$ is large.

While multi-threading on CPUs can initiate multiple walkers in parallel, it may not scale well for large $s$ or large batches of updates. In contrast, the SIMT (Single Instruction, Multiple Threads) architecture of modern GPUs is inherently well-suited to random-walk parallelization~\cite{C_SAW}. By assigning each GPU thread a walker, we can process many edges (or many walks) in parallel, significantly reducing the wall clock time for each batch of updates.

However, supporting dynamic updates efficiently on the GPU presents two key challenges:
(1) \textbf{fast neighbor queries} during the walk, and 
(2) \textbf{fast edge insertions/deletions} for both incremental and decremental updates. 
Traditional data structures like the Compressed Sparse Row (CSR) format enable efficient neighbor access in $O(1)$ time but require costly full reconstruction upon edge insertions or deletions. In contrast, the Coordinate (COO) format supports more flexible edge updates but suffers from inefficient neighbor queries due to its unsorted and unindexed layout.
To address these issues, recent GPU frameworks have proposed more general-purpose dynamic graph data structures optimized for high-throughput workloads. For example, LPMA~\cite{zou2023efficient} improves update performance through a leveled tree layout and localized re-balancing strategies, while a hash-based design~\cite{awad2020dynamic} assigns each vertex a dedicated hash table to enable constant-time edge insertions and deletions. These systems achieve impressive scalability and support large-scale dynamic workloads efficiently.
However, these solutions are primarily optimized for high-churn settings where frequent large batches of updates dominate the workload. In our application, edge updates are relatively moderate—typically around 25\% of the vertex count—and query performance is more critical than raw update speed. 

\begin{figure}[t]
    \centering
    \includegraphics[width=0.4305\textwidth]{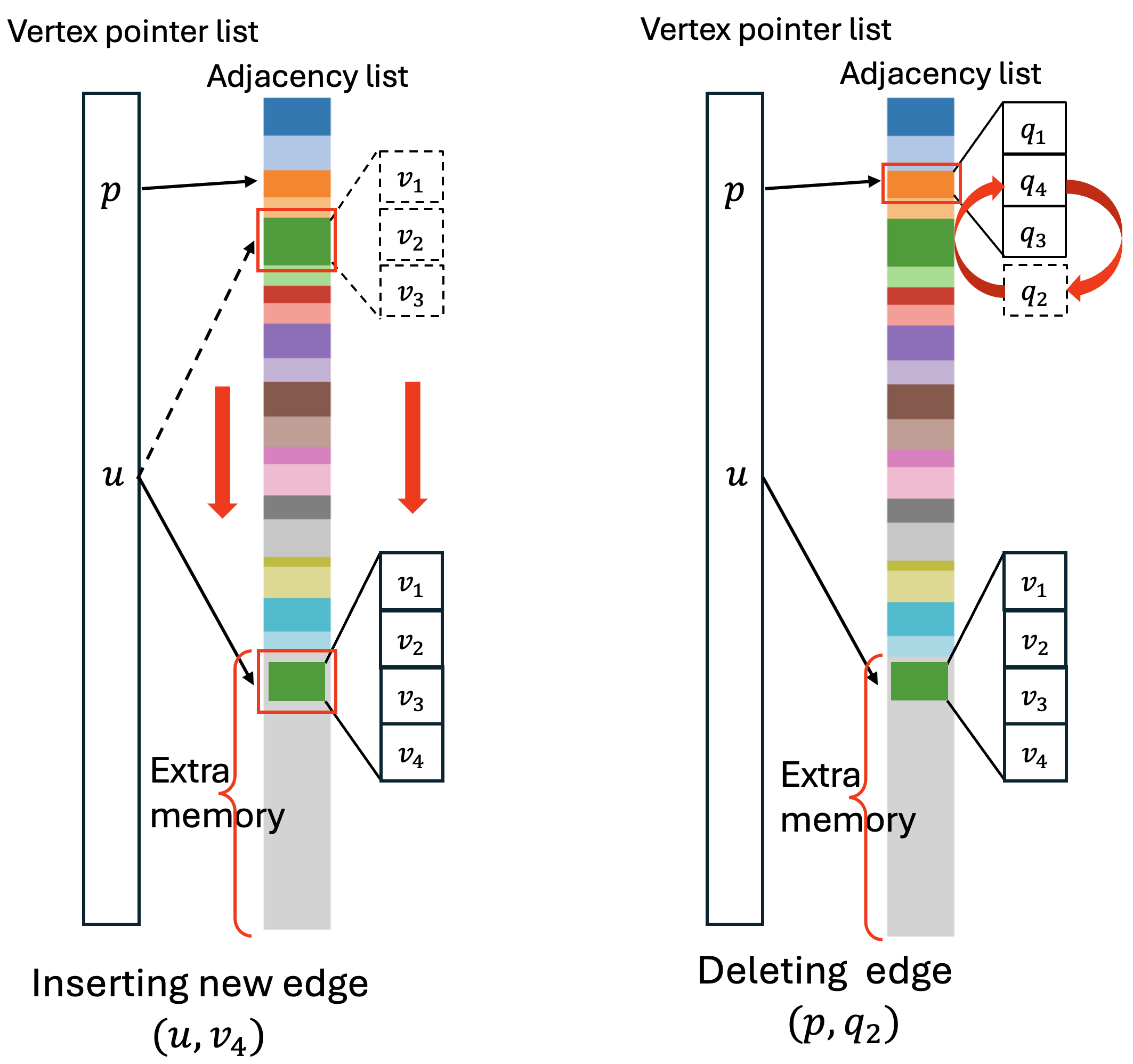}
    \caption{Schematic of dynamic CSR data structure illustrating insertion of $(u,v_4)$ and deletion of $(p,q_2)$.}
    \label{figure:dyn_CSR}
\end{figure}

To this end, we design a lightweight, self-defined dynamic CSR structure that replaces traditional row offsets with a \emph{vertex pointer list} that directly references each vertex’s \emph{adjacency lists} in global memory. During initialization, we allocate an extra memory block to accommodate future edge insertions. As shown in Fig.~\ref{figure:dyn_CSR}, incremental updates write new neighbor sub-arrays into this extra space and update the pointers in the vertex pointer array accordingly. Decremental updates, by contrast, can be handled in-place without additional memory overhead. This design preserves efficient neighbor lookups while permitting updates without reconstructing the entire data structure.
In the GPU setting, we adopt a batch-processing model to maximize throughput and reduce host-device synchronization. Unlike the CPU version of \textit{dyGRASS}, which applies updates immediately, the GPU version delays updates until the end of a batch. This deferral may slightly affect sparsifier quality in the short term, but it substantially improves overall efficiency by minimizing PCIe traffic and increasing GPU utilization.
To parallelize random walks, we assign one GPU thread per walker. Each thread performs a non-backtracking random walk independently, using the dynamically updated adjacency array. Within a thread block, all walkers handle the same edge insertion or deletion in parallel. At each step, a thread reads the degree of its current vertex, follows the corresponding pointer in the vertex pointer array to retrieve the neighbor sub-array, and selects a neighbor other than the one visited in the previous step—thereby enforcing the non-backtracking constraint.
This lightweight approach balances dynamic updates with fast neighborhood exploration, making it a practical choice for GPU-accelerated random walks in \textit{dyGRASS}.

\section{Experimental results}\label{sec:result}

This section outlines the findings from a series of experiments designed to evaluate the effectiveness and efficiency of the proposed \textit{dyGRASS}. These experiments are conducted using a variety of public domain sparse matrices \footnote{available at https://sparse.tamu.edu/} originating from various real-world applications, such as circuit simulation and finite element analysis. All experiments were conducted on a Linux Ubuntu $24.04$ system with $64$ GB of RAM, a $3.4$ GHz $8$-core AMD R7 5800X3D CPU, and an Nvidia RTX 4090 GPU. The state-of-the-art spectral sparsification tool \textit{GRASS} \cite{feng2020grass} \footnote{https://sites.google.com/mtu.edu/zhuofeng-graphspar/home} has been used to generate the initial sparsifier, while the state-of-the-art  incremental spectral sparsification tool \textit{inGRASS} \cite{aghdaei2024ingrass} served as the baseline for assessing the performance and scalability of \textit{dyGRASS}.

\begin{table}[]
\caption{Graph density and spectral similarity of the initial (updated) spectral sparsifiers $H^{(0)}$ ($H^{(10)}$).}
\centering
\begin{tabular}{|c|c|c|c|c|}
\hline
\multicolumn{1}{|c|}{Test Cases} & $|V|$  & $|E|$  & $d_H^{(0)} \rightarrow d_H^{(10)}$& \begin{tabular}[c]{@{}l@{}}$\kappa (L_G^{(0)},L_H^{(0)})$\\
$\rightarrow$ \\
$\kappa (L_G^{(10)},L_H^{(0)})$\end{tabular}\\ \hline
G2\_circuit                      & 1.5E+5 & 3.0E+6 & 10.0$\%\rightarrow$32.3\% & 71.9$\rightarrow$397.5            \\ \hline
G3\_circuit                      & 1.5E+6 & 2.9E+5 & 10.5$\%\rightarrow$33.8\% & 88.4$\rightarrow$349.42            \\ \hline
fe\_4elt                         & 1.1E+4 & 3.3E+4 & 9.9$\%\rightarrow$40.1\%  & 95.8$\rightarrow$317.65            \\ \hline
fe\_ocean                        & 1.4E+5 & 4.1E+5 & 9.8$\%\rightarrow$39.6\%  & 212.5$\rightarrow$481.00           \\ \hline
fe\_sphere                       & 1.6E+4 & 4.9E+4 & 10.5$\%\rightarrow$35.3\% & 124.9$\rightarrow$782.69          \\ \hline
delaunay\_n18                    & 2.6E+5 & 6.5E+5 & 10.5$\%\rightarrow$34.8\% & 115.9$\rightarrow$312.01           \\ \hline
delaunay\_n19                    & 5.2E+5 & 1.6E+6 & 10.6$\%\rightarrow$35.2\% & 124.1$\rightarrow$371.08           \\ \hline
delaunay\_n20                    & 1.0E+6 & 3.1E+6 & 10.5$\%\rightarrow$35.0\% & 129.0$\rightarrow$391.59           \\ \hline
delaunay\_n21                    & 2.1E+6 & 6.3E+6 & 10.1$\%\rightarrow$34.6\% & 153.2$\rightarrow$400.46           \\ \hline
delaunay\_n22                    & 4.2E+6 & 1.3E+7 & 10.3$\%\rightarrow$34.8\% & 164.9$\rightarrow$429.81           \\ \hline
M6                               & 3.5E+6 & 1.1E+7 & 9.8$\%\rightarrow$33.9\%  & 174.0$\rightarrow$777.39          \\ \hline
333SP                            & 3.7E+6 & 1.1E+7 & 9.7$\%\rightarrow$33.7\%  & 182.6$\rightarrow$1067.05          \\ \hline
AS365                            & 3.8E+6 & 1.1E+7 & 10.1$\%\rightarrow$34.2\% & 164.8$\rightarrow$1841.68          \\ \hline
NACA15                           & 1.0E+6 & 3.1E+6 & 10.5$\%\rightarrow$34.5\% & 150.8$\rightarrow$526.55          \\ \hline
\end{tabular}
\label{tab:setup}
\end{table}

\begin{table*}[]
\caption{Comparison of Dynamic Graph Spectral Sparsification Outcomes through $10$-Iterative Updates applying i\textit{inGRASS} and \textit{dyGRASS}.}
\centering
\begin{tabular}{|c|cccc|ccc|ccc|}
\hline
 & \multicolumn{4}{c|}{inGRASS.incremental() \cite{aghdaei2024ingrass} } & \multicolumn{3}{c|}{dyGRASS.incremental()} & \multicolumn{3}{c|}{dyGRASS.decremental()} \\ \cline{2-11} 
\multirow{-2}{*}{Test Cases} & \multicolumn{1}{c|}{$T_{setup}$} & \multicolumn{1}{c|}{$T_{update}$} & \multicolumn{1}{c|}{$d_H^{(10)}$} & $\kappa (L_G^{(10)},L_H^{(10)})$ & \multicolumn{1}{c|}{$T_{update}$} & \multicolumn{1}{c|}{$d_H^{(10)}$} & $\kappa (L_G^{(10)},L_H^{(10)})$ & \multicolumn{1}{c|}{$T_{update}$} & \multicolumn{1}{c|}{$d_H^{(20)}$} & $\kappa (L_G^{(20)},L_H^{(20)})$ \\ \hline
G2\_circuit                  & \multicolumn{1}{c|}{1.77s}        & \multicolumn{1}{c|}{0.01s}         & \multicolumn{1}{c|}{{\color[HTML]{333333} 13.97\%}} & 95.41                                                       & \multicolumn{1}{c|}{0.019s}    & \multicolumn{1}{c|}{{\color[HTML]{333333} \textbf{12.10\%}}} & \textbf{71.91}  & \multicolumn{1}{c|}{0.022s}    & \multicolumn{1}{c|}{12.87\%}            & 70.15                                                       \\ \hline
G3\_circuit                  & \multicolumn{1}{c|}{23.5s}        & \multicolumn{1}{c|}{1.20s}         & \multicolumn{1}{c|}{{\color[HTML]{333333} 13.91\%}} & \textbf{88.35}                                                       & \multicolumn{1}{c|}{0.088s}    & \multicolumn{1}{c|}{{\color[HTML]{333333} \textbf{13.58\%}}} & 88.39           & \multicolumn{1}{c|}{0.155s}    & \multicolumn{1}{c|}{14.37\%}            & 113.51                                                      \\ \hline
fe\_4elt                     & \multicolumn{1}{c|}{0.12s}        & \multicolumn{1}{c|}{0.01s}         & \multicolumn{1}{c|}{{\color[HTML]{333333} 15.01\%}} & 136.32                                                      & \multicolumn{1}{c|}{0.010s}    & \multicolumn{1}{c|}{{\color[HTML]{333333} \textbf{13.87\%}}} & \textbf{98.59}  & \multicolumn{1}{c|}{0.011s}    & \multicolumn{1}{c|}{14.40\%}            & 98.54                                                       \\ \hline
fe\_ocean                    & \multicolumn{1}{c|}{1.77s}        & \multicolumn{1}{c|}{0.11s}         & \multicolumn{1}{c|}{{\color[HTML]{333333} 16.34\%}} & 206.97                                                      & \multicolumn{1}{c|}{0.040s}    & \multicolumn{1}{c|}{{\color[HTML]{333333} \textbf{13.39\%}}} & \textbf{209.70} & \multicolumn{1}{c|}{0.026s}    & \multicolumn{1}{c|}{14.29\%}            & 194.92                                                      \\ \hline
fe\_sphere                   & \multicolumn{1}{c|}{0.16s}        & \multicolumn{1}{c|}{0.01s}         & \multicolumn{1}{c|}{{\color[HTML]{333333} 14.07\%}} & 173.35                                                      & \multicolumn{1}{c|}{0.010s}    & \multicolumn{1}{c|}{{\color[HTML]{333333} \textbf{13.88\%}}} & \textbf{144.77} & \multicolumn{1}{c|}{0.011s}    & \multicolumn{1}{c|}{14.41\%}            & 144.68                                                      \\ \hline
delaunay\_n18                & \multicolumn{1}{c|}{3.57s}        & \multicolumn{1}{c|}{0.20s}         & \multicolumn{1}{c|}{{\color[HTML]{333333} 15.29\%}} & 171.04                                                      & \multicolumn{1}{c|}{0.024s}    & \multicolumn{1}{c|}{{\color[HTML]{333333} \textbf{14.56\%}}} & \textbf{132.77} & \multicolumn{1}{c|}{0.038s}    & \multicolumn{1}{c|}{15.05\%}            & 132.70                                                      \\ \hline
delaunay\_n19                & \multicolumn{1}{c|}{7.79s}        & \multicolumn{1}{c|}{0.39s}         & \multicolumn{1}{c|}{{\color[HTML]{333333} 18.73\%}} & 138.18                                                      & \multicolumn{1}{c|}{0.040s}    & \multicolumn{1}{c|}{{\color[HTML]{333333} \textbf{15.44\%}}} & \textbf{127.23} & \multicolumn{1}{c|}{0.069s}    & \multicolumn{1}{c|}{15.92\%}            & 129.07                                                      \\ \hline
delaunay\_n20                & \multicolumn{1}{c|}{17.02s}       & \multicolumn{1}{c|}{0.75s}         & \multicolumn{1}{c|}{{\color[HTML]{333333} 18.43\%}} & 148.13                                                      & \multicolumn{1}{c|}{0.075s}    & \multicolumn{1}{c|}{{\color[HTML]{333333} \textbf{15.09\%}}} & \textbf{146.61} & \multicolumn{1}{c|}{0.127s}    & \multicolumn{1}{c|}{15.58\%}            & 146.60                                                      \\ \hline
delaunay\_n21                & \multicolumn{1}{c|}{36.69s}       & \multicolumn{1}{c|}{1.60s}         & \multicolumn{1}{c|}{{\color[HTML]{333333} 18.00\%}} & 159.90                                                      & \multicolumn{1}{c|}{0.145s}    & \multicolumn{1}{c|}{{\color[HTML]{333333} \textbf{14.50\%}}} & \textbf{147.54} & \multicolumn{1}{c|}{0.273s}    & \multicolumn{1}{c|}{14.99\%}            & 147.10                                                      \\ \hline
delaunay\_n22                & \multicolumn{1}{c|}{80.26s}       & \multicolumn{1}{c|}{4.32s}         & \multicolumn{1}{c|}{{\color[HTML]{333333} 18.19\%}} & 176.14                                                      & \multicolumn{1}{c|}{0.283s}    & \multicolumn{1}{c|}{{\color[HTML]{333333} \textbf{14.40\%}}} & \textbf{164.36} & \multicolumn{1}{c|}{0.497s}    & \multicolumn{1}{c|}{14.90\%}            & 164.37                                                      \\ \hline
M6                           & \multicolumn{1}{c|}{66.53s}       & \multicolumn{1}{c|}{3.72s}         & \multicolumn{1}{c|}{{\color[HTML]{333333} 17.73\%}} & 232.50                                                      & \multicolumn{1}{c|}{0.235s}    & \multicolumn{1}{c|}{{\color[HTML]{333333} \textbf{13.43\%}}} & \textbf{173.64} & \multicolumn{1}{c|}{0.474s}    & \multicolumn{1}{c|}{13.92\%}            & 173.63                                                      \\ \hline
333SP                        & \multicolumn{1}{c|}{67.00s}       & \multicolumn{1}{c|}{3.68s}         & \multicolumn{1}{c|}{{\color[HTML]{333333} 14.13\%}} & 232.57                                                      & \multicolumn{1}{c|}{0.251s}    & \multicolumn{1}{c|}{{\color[HTML]{333333} \textbf{13.61\%}}} & \textbf{182.46} & \multicolumn{1}{c|}{0.475s}    & \multicolumn{1}{c|}{14.11\%}            & 183.39                                                      \\ \hline
AS365                        & \multicolumn{1}{c|}{73.26s}       & \multicolumn{1}{c|}{4.30s}         & \multicolumn{1}{c|}{{\color[HTML]{333333} 14.63\%}} & 196.76                                                      & \multicolumn{1}{c|}{0.258s}    & \multicolumn{1}{c|}{{\color[HTML]{333333} \textbf{13.6\%}}}  & \textbf{183.64} & \multicolumn{1}{c|}{0.490s}    & \multicolumn{1}{c|}{14.10\%}            & 183.33                                                      \\ \hline
NACA15                       & \multicolumn{1}{c|}{17.14s}       & \multicolumn{1}{c|}{1.00s}         & \multicolumn{1}{c|}{{\color[HTML]{333333} 15.19\%}} & 219.66                                                      & \multicolumn{1}{c|}{0.074s}    & \multicolumn{1}{c|}{{\color[HTML]{333333} \textbf{13.94\%}}} & \textbf{150.16} & \multicolumn{1}{c|}{0.137s}    & \multicolumn{1}{c|}{14.44\%}            & 150.17                                                      \\ \hline
\end{tabular}
\label{tab:main}
\end{table*}

\subsection{Dynamic spectral sparsification using \textit{dyGRASS}}

Consistent with the prior \textit{inGRASS} \cite{aghdaei2024ingrass} study, we reuse its 14 benchmark graphs and replay the identical update stream: ten insertion batches that together add $\approx\!25\%$ of $|V|$ edges. We evaluate the performance of \textit{dyGRASS} by comparing the off-tree edge density of the sparsifier, which is defined as $d_H:= \frac{|E|}{|V|} - 1$  the same relative condition number is achieved using \textit{inGRASS} and \textit{dyGRASS} methods. The relative condition number $\kappa (L_G,L_H)$ can be deployed to quantify the spectral similarity between the original graph $G$ and its graph sparsifier $H$ \cite{feng2016spectral,feng2020grass}. A smaller relative condition number implies higher spectral similarity between the graph $G$ and its sparsifier $H$. 

 As shown in Table \ref{tab:setup}, the  graph sparsifier density  increases significantly ($d_H^{(0)}\rightarrow$ $d_H^{(10)}$) when all the edges are inserted into $H^{(0)}$ through $10$-iterative updates  \textbf{without} an edge filtering mechanism. On the other hand, if we keep the initial graph sparsifier $H^{(0)}$   unchanged through $10$-iterative updates, the relative condition number $\kappa (L_G^{(10)},L_H^{(0)})$ will grow significantly from the initial condition number $\kappa (L_G^{(0)},L_H^{(0)})$, implying increasing mismatches between graph $G^{(10)}$ and the initial sparsifier $H^{(0)}$.

Table \ref{tab:main} compares GPU-accelerated \textit{dyGRASS}   and \textit{inGRASS} \cite{aghdaei2024ingrass} over $10$ incremental updates. During the   ten incremental updates, approximately $25\sim 30\%$ edges in total are inserted into $G^{(0)}$, as shown in column $4$ of Table \ref{tab:setup}. Subsequently, we also perform ten decremental updates using \textit{dyGRASS}, in which $1\%$ edges in total are randomly removed from $G^{(10)}$. Each incremental or decremental update is configured to produce sparsifiers matching the original spectral similarity $\kappa(L_G^{(0)},L_H^{(0)})$. As shown in Table \ref{tab:main},  dyGRASS.incremental() consistently achieves   lower or similar sparsifier densities compared to ones produced by inGRASS.incremental(). Additionally, for the subsequent ten decremental updates, dyGRASS.decremental() is always able to efficiently preserve   low densities of the updated spectral sparsifiers $(d_H^{(20)})$ while preserving high spectral quality. 

\vspace{-5pt}

\subsection{Sparsifier densities after multiple incremental updates}
%.8599895
\begin{figure}
    \centering
    \includegraphics [width = .80\linewidth]{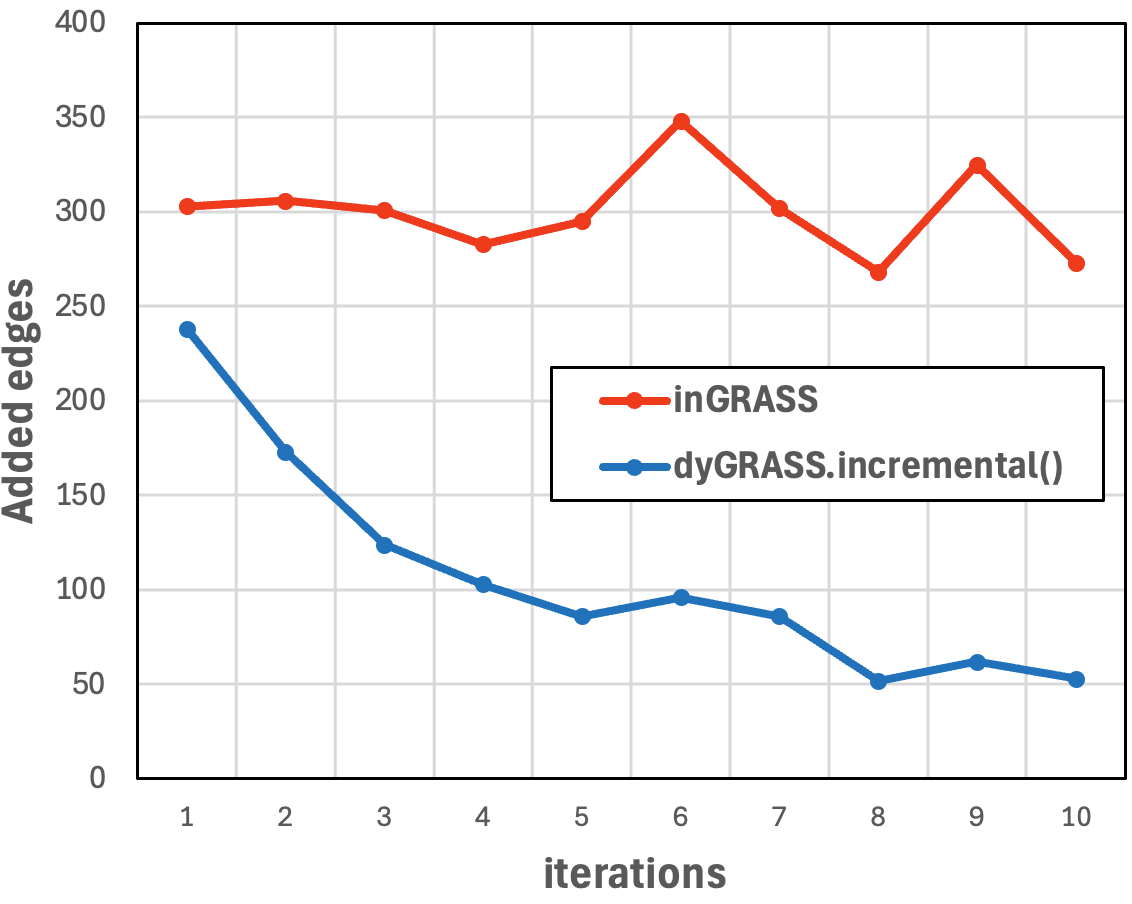}
    \caption{Comparison of sparsifier densities in 10 incremental updates}
    \label{fig:edge_insertion}
\end{figure}
Fig. \ref{fig:edge_insertion} illustrates \textit{dyGRASS}’s dynamic updating capability in ten iterations for handling the circuit graph “G2\_circuit”. Since \textit{dyGRASS} can capture the latest structural changes (e.g., decreasing resistance diameters) in every update iteration via localized random walks, it allows the newly inserted edges to be selected more optimally considering the most recent updates. As a result, \textit{dyGRASS} is able to achieve much lower sparsifier density than \textit{inGRASS} as more updates are made.  In contrast,   when more   edges are inserted into the graph, the resistance distances between nodes will monotonically decrease, leading to an increasing overestimation of the spectral distortion of each new edge since \textit{inGRASS} keeps processing newly inserted edges using the dated resistance embedding.

\renewcommand{\arraystretch}{1.2}
\begin{table}[]
\centering
\caption{Solver Iterations and Sparsifier Densities with/without dyGRASS Updates.}
\setlength{\tabcolsep}{4pt}
\begin{tabular}{|lll|ll|ll|}

\hline
\multicolumn{3}{|c|}{Initial sparsifier}                                                                                           & \multicolumn{2}{c|}{No update}                                                   & \multicolumn{2}{c|}{dyGRASS update}                                            \\ \hline
\multicolumn{1}{|c|}{$\kappa (L_G^{(0)},L_H^{(0)})$} & \multicolumn{1}{c|}{\textbf{$d_H^{(0)}$}} & \multicolumn{1}{c|}{\#Iters} & \multicolumn{1}{c|}{\textbf{ $\Delta d_G^{(10)}$ }} & \multicolumn{1}{c|}{\#Iters} & \multicolumn{1}{c|}{\textbf{$d_H^{(10)}$}} & \multicolumn{1}{c|}{\#Iters} \\ \hline
\multicolumn{1}{|c|}{107.69}       & \multicolumn{1}{c|}{13.77\%}                         & 94                  & \multicolumn{1}{c|}{24.10\%}                               & 273                 & \multicolumn{1}{c|}{17.4\%}                          & 103                 \\ \hline
\multicolumn{1}{|c|}{183.18}       & \multicolumn{1}{c|}{8.52\%}                          & 122                 & \multicolumn{1}{c|}{24.10\%}                               & 336                 & \multicolumn{1}{c|}{12.33\%}                         & 120                 \\ \hline
\multicolumn{1}{|c|}{407.92}       & \multicolumn{1}{c|}{4.92\%}                          & 171                 & \multicolumn{1}{c|}{24.10\%}                               & 426                 & \multicolumn{1}{c|}{9.64\%}                          & 152                 \\ \hline
\end{tabular}

\label{tab:solver_iterations}
\end{table}

\subsection{Preconditioned Solver Iterations}
To solve a linear system $L_G \,x = b$ more efficiently, we use a spectral sparsifier $H$ as a preconditioner. That is, we replace $L_G$ by $L_H$ in each iterative step, effectively solving $L_H^{-1} L_G \,x = L_H^{-1} b$. The convergence rate hinges on the condition number $\kappa(L_G, L_H)$, so a high-quality sparsifier leads to fewer solver iterations. In this experiemnt, the PCG solver terminates when the relative residual norm $<10^{-8}$.

Table~\ref{tab:solver_iterations} considers three initial sparsifiers of the “AS365” graph, each with a different condition number and corresponding iteration count (\#Iters). After inserting $24.1\%$ new edges into $G^{(0)}$ to form $G^{(10)}$, leaving the preconditioner $H^{(0)}$ unchanged (\textbf{No update}) inflates the iteration count significantly. In contrast, applying \textit{dyGRASS} updates (\textbf{dyGRASS update}) slightly increases the off-tree density but drastically reduces the iteration count. This demonstrates how \textit{dyGRASS} retains the crucial edges for preconditioning while pruning less important ones, maintaining a low-density sparsifier that preserves robust solver performance despite added edges.

\subsection{Runtime scalability of \textit{dyGRASS}}
\begin{figure}
    \centering
    \includegraphics [width = .80\linewidth]{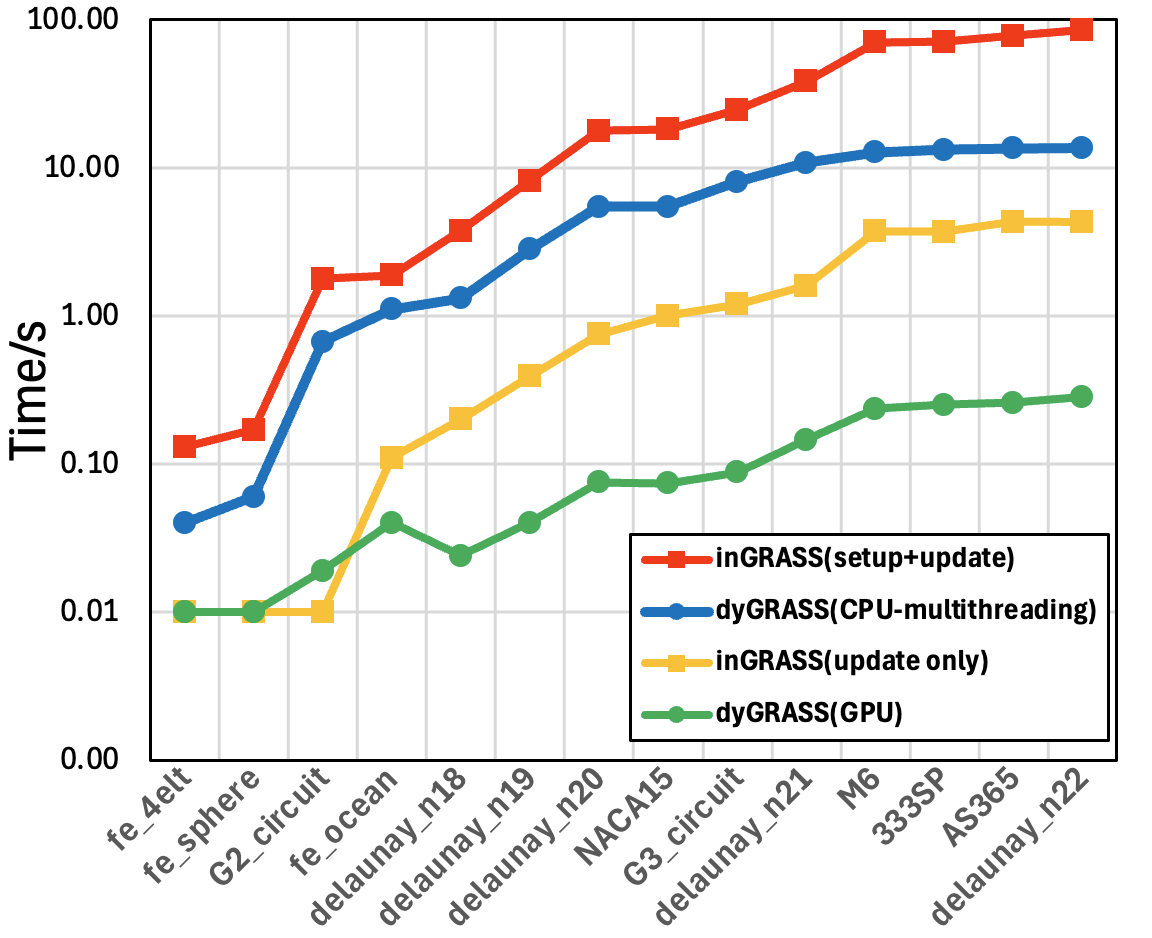}
    \caption{Runtime scalability comparison between \textit{inGRASS} and \textit{dyGRASS}.}
    \label{fig:runtime}
\end{figure}

In Fig.~\ref{fig:runtime}, we compare the runtime scalability of \textit{dyGRASS} and \textit{inGRASS} across various real-world graphs. Over ten iterative incremental updates, the GPU implementation of \textit{dyGRASS} achieves approximately a 70$\times$ speedup over its CPU counterpart. Compared to \textit{inGRASS}, \textit{dyGRASS} offers about a 10$\times$ improvement for large graphs.
Although \textit{dyGRASS} maintains $O(1)$ time complexity per edge update, the total runtime increases with graph size due to the growing number of updates. Specifically, we fix the update density at approximately 25\% of the number of vertices. As a result, larger graphs involve proportionally more edge updates per iteration, leading to an increase in total update time.

\vspace{-10pt}

\section{Conclusion}\label{sec:conclusion}
This work presents dyGRASS, an efficient and scalable algorithm for dynamic spectral sparsification of large undirected graphs. By employing random-walk-based techniques for incremental and decremental updates, dyGRASS can efficiently identify spectrally critical edges essential for preserving graph spectral properties while pruning redundant edges to maintain desired sparsifier density. The proposed dyGRASS algorithm also leverages GPU acceleration to efficiently process a large number of dynamic updates. Our extensive experimental results demonstrate that dyGRASS outperforms state-of-the-art methods, \textit{inGRASS}, in both runtime and solution quality. It achieves significant speedups while dynamically maintaining spectral fidelity across various graph types, including those from circuit simulations, finite element analysis, and social networks. Overall, the \textit{dyGRASS} framework bridges the gap between theoretical advancements in spectral graph theory and practical EDA applications, providing a robust and scalable solution for compressing dynamically evolving graphs.

\section{Acknowledgments}
This work is supported in part by the National Science
Foundation under Grants CCF-2417619 and CCF-2212370.

\bibliographystyle{IEEEtran}
\bibliography{Ref,feng}

\end{document}